\documentclass[nofootinbib,preprintnumbers,amsmath,amssymb,twocolumn,unsortedaddress,superscriptaddress]{revtex4-1}
\usepackage[a4paper, total={6.5in, 9.7in}]{geometry}
\usepackage{setspace}

\usepackage[utf8]{inputenc}
\usepackage{dcolumn}
\usepackage{rotating}
\usepackage[table]{xcolor}
\usepackage{abraces}
\usepackage{siunitx}
\usepackage{hyperref}
\usepackage{float}
\usepackage{multirow}
\usepackage{mathtools}
\usepackage{bm}
\usepackage{cleveref}
\usepackage[normalem]{ulem}
\usepackage{soul}

\newcommand{\REV}[1]{\textcolor{black}{{#1}}} 
\renewcommand{\vec}[1]{\bm{ #1 }}

\definecolor{alat}{HTML}{e377c2}
\definecolor{clat}{HTML}{d36b69}
\definecolor{covalat}{HTML}{c15d00}
\definecolor{zatom}{HTML}{984ea3}
\newcommand{\alattice}[1]{\textsc{\textbf{\textit{\textcolor{alat}{{#1}}}}}} 
\newcommand{\clattice}[1]{\textsc{\textbf{\textit{\textcolor{clat}{{#1}}}}}} 
\newcommand{\zz}[1]{\textsc{\textbf{\textcolor{zatom}{{#1}}}}} 
\newcolumntype{C}{>{\centering\arraybackslash}m{1.20in}}
\newcolumntype{T}{>{\centering\arraybackslash}m{0.7in}}
\begin{document}

\title{Parametrically Constrained Geometry Relaxations for High-Throughput Materials Science}
\author{Maja-Olivia Lenz}
\author{Thomas A. R. Purcell*}
\affiliation{Fritz-Haber-Institut der Max-Planck-Gesellschaft, Faradayweg 4–6, D-14195 Berlin, Germany}
\author{David Hicks}
\author{Stefano Curtarolo}
\affiliation{Department of Materials Science and Mechanical Engineering, Duke University, Durham, NC 27708, USA}
\author{Matthias Scheffler}
\author{Christian Carbogno}
\affiliation{Fritz-Haber-Institut der Max-Planck-Gesellschaft, Faradayweg 4–6, D-14195 Berlin, Germany}
\date{\today}

\begin{abstract}
Reducing parameter spaces via exploiting symmetries has greatly accelerated and increased the quality of electronic-structure calculations.
Unfortunately, many of the traditional methods fail when the global crystal symmetry is broken, even when the distortion is only a slight perturbation (e.g. Jahn-Teller like distortions).
Here we introduce a flexible \REV{and generalizable} parametric relaxation scheme, and implement it in the all-electron code FHI-aims.
This approach utilizes \REV{parametric} constraints to maintain symmetry at any level.
After demonstrating the method's ability to relax metastable structures, we highlight its adaptability and performance over a test set of \REV{359} materials, across thirteen lattice prototypes.
Finally we show how these constraints can reduce the number of steps needed to relax local lattice distortions by an order of magnitude.
The flexibility of these constraints enables a significant acceleration of the high-throughput searches for novel materials \REV{for numerous applications}.

\end{abstract}

\maketitle
\section{Introduction}
Symmetry preservation and breaking is one of the most fundamental processes in physics and chemistry.
Many properties and applications such as,~piezoelectricity~\cite{Panda2009,Wang2012,Ok2016,Nguyen2013a}, pyroelectricity~\cite{Moure2015a,Isaenko2015a},~ferroelectricity~\cite{Damjanovic1998,Hang2011a,Sun2014c,Shi2016a},~topological insulators~\cite{Cava2013,Fiete2012a}, and non-linear optics~\cite{Evans2002a,Wu2018a,Xu2019a}, require certain selection rules to be met, and therefore require certain crystallographic symmetries to be maintained.
Furthermore, it is not only {\bf global} symmetry, the space and point groups of a material, but also {\bf local} symmetry breaking that matters.
For example, defects can cause significant changes in a material's mechanical~\cite{Le2014, Sayle} and optical~\cite{Pan,Wang2016d} properties as well as in its electronic~\cite{An2014,Anno1999,Senanayak2017} and thermal transport~\cite{Proshchenko2019,Toberer2010,Xie2013} coefficients.
\REV{These effects can be particularly important at thin-film interfaces where interactions between different layers can induce systematic distortions in the structure of the film.
In turn, these distortions can lead to novel properties in the materials, such as artificial ferroelectricity in layered perovskite supercell structures~\cite{Wang2016,Bousquet2008}.
Clearly, such problems cannot be addressed by enforcing a \textbf{global} symmetry constraint,~e.g.,~space group conservation, for the whole system, but require to
selectively preserve and break material's symmetries \textbf{locally},~e.g.,~around defects or in the individual perovskite layers. This is paramount in computational material science, especially in {\it high-throughput} studies, which often aim at calculating and exploring yet unknown properties of already known materials,~e.g.,~band structures~\cite{Ricci2017}, defect formation energies~\cite{Broberg2018},  elastic and thermal properties~\cite{Plata2017a} and topological constants~\cite{Tang2019,Maghirang2019}. Similarly, many high-throughput studies aim at discovering potentially stable or meta-stable materials by decorating complex, well-known crystal structures such as Hauslers~\cite{Roy} and perovskites~\cite{Mazaheri2019} with different species, or by systematically exploring a given alloy system~\cite{Sutton2018}. To streamline such calculations it is essential to keep both global and local symmetries under control, especially when complex materials or material properties are targeted.
In this work, we achieve this goal by proposing and implementing parametric geometric constraints that allow for enforcing or breaking symmetries both globally and locally. Before applying these constraints, an understanding of how the constraints map onto the target systems (e.g.,~the number of atoms in the unit cell, space group, or the effects of local distortions on the structure) is necessary. To facilitate setting up such constraints, we rely on the AFLOW Library of Crystallographic Prototypes~\cite{Mehl2017, Hicks2019} to generate the initial mapping of real space onto a reduced parameter space that fully describes a system. One can then manually alter the initial mapping to add or lift constraints as needed. This allows for the efficient targeting of specific geometric configurations and avoids revisiting and recalculating already investigated configurations.
}

Traditionally, crystallographic symmetries are incorporated in first-principles codes already at the electronic-structure level~(e.g.,~by sampling $\vec{k}$-space grids in the irreducible part of the Brillouin zone\REV{, as implemented in Vasp}~\cite{Hafner2008} or by sampling real space in symmetry defined ``irreducible wedges''\REV{, as done in parsec}~\cite{Kronik2006}) since it leads to significant savings in memory and computational workload for highly symmetric crystals.
Also, by this means the obtained forces on the atoms and stresses on the lattice vectors fully reflect the crystallographic symmetries.
Since geometry relaxation algorithms such as steepest descent, conjugate gradient, Newton-Raphson, quasi-Newton (e.g. BFGS~\cite{Fletcher1987}), and truncated-Newton methods~\cite{NoceWrig06} rely on the forces and stresses to update the atomic and lattice degrees of freedom, {\bf global} symmetries are inherently preserved in such approaches.
However, this does not allow for the {\bf \REV{partial}}, local symmetry breaking discussed in the introduction. To address such cases in first-principles calculations, it is typically necessary to lift  all crystallographic symmetry constraints and treat the atomic and lattice degrees of freedom as a set of freely changing parameters.
Besides the increased computational cost, such unconstrained structure optimization can lead to long and inefficient relaxation trajectories\REV{, resulting in structures far from the ideal geometry}.
\REV{While in some cases this problem can be circumvented by fixing atomic, lattice, or internal~\cite{Freysoldt2017, Panosetti2015} degrees of freedom (as done in Quantum Espresso~\cite{Giannozzi2009} or Vasp~\cite{Hafner2008}), mapping local distortions onto these constraints requires cumbersome manual inspection and analysis, if it is even possible.
A more direct approach targeting how the distortion changes the native crystal structure provides an easier and better way of treating these systems.
}

Here we present a scheme to incorporate parametric constraints in structure optimizations that treats all levels of symmetry equally.
The proposed approach employs a mapping of the relevant degrees of freedom onto a lower-dimensional representation of the structure; the respective forces and stresses are then automatically mapped in this reduced representation.
With that, the implemented formalism does not require to alter the employed relaxation algorithm, while still allowing the introduction of arbitrary constraints in a user-friendly manner.
We first describe how the methodology works and the tools that can be used to quickly generate new structures.
We demonstrate that these constraints \REV{allow for performing} geometry optimizations on dynamically stabilized structures, which are not easily addressable otherwise.
By analyzing the constrained and unconstrained relaxations of a test set of \REV{359} materials, we then show that these constraints are also computationally beneficial for the relaxation of stable materials.
Finally we illustrate how the parameters can be used to \REV{{\bf selectively break symmetries}} and accelerate relaxations in supercells.

\section{Transformation to reduced space}
\label{sec:transformation}
For a free relaxation, the optimizer acts on the full $3N+9$ dimensional potential-energy surface~$E(\vec{R},\vec{L})$ of a material, which is encoded by the atomic,~$\vec{R}$, and lattice,~$\vec{L}$, degrees of freedom.
The lattice degrees of freedom are stored as the three components of the three lattice vectors in the chosen unit cell, and the atomic degrees of freedom are the $3N$ \REV{components of the} positions of the $N$ atoms in a unit cell, represented by Cartesian or fractional coordinates.
The forces, $\vec{F}$, acting on the atomic degrees of freedom are the derivatives \REV{of} the energy with respect to $\vec{R}$
\begin{equation}
    \vec{F}=-\frac{dE}{d\vec{R}},
\end{equation}
while the forces acting on the lattice vectors stem from the stresses,~$\sigma$,
\begin{equation}
    \sigma =\frac{1}{V}\frac{dE}{d\vec{L}},
\end{equation}
where $V$ is the volume of the unit cell.
In {\it ab initio} approaches, $E$ is determined by solving the electronic-structure problem, and the respective derivatives are obtained analytically via the Hellmann-Feynman Theorem.
However, in practice this requires to account for additional terms, such as the Pulay terms and multipole corrections, as done in FHI-aims~\cite{Blum2009,Knuth2015}.

\begin{figure*}
    \centering
    \includegraphics{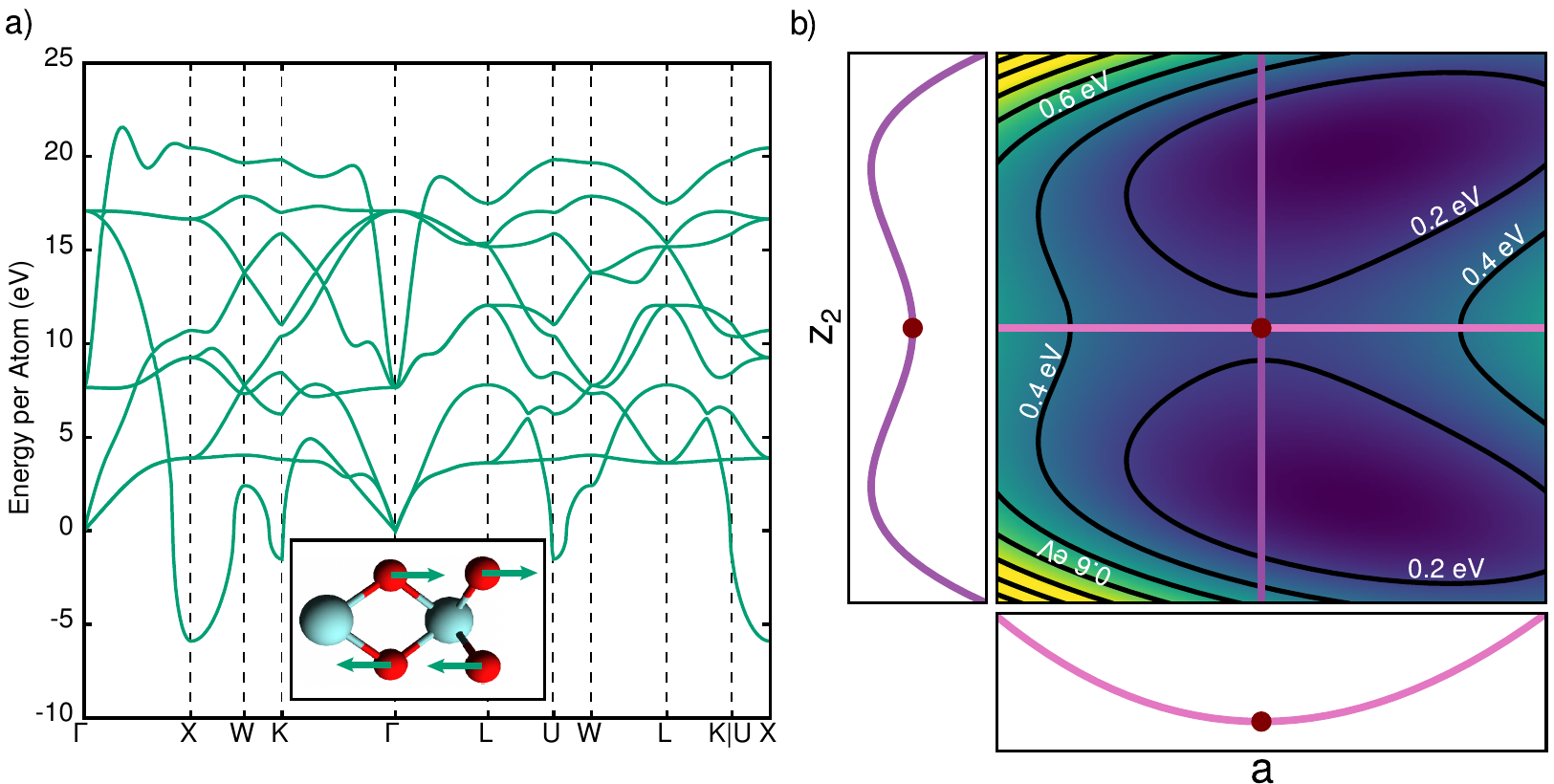}
    \caption{An illustration of the new relaxation scheme. a) \REV{The phonon band structures of cubic ZrO$_2$. The inset illustrates the eigenvector of the imaginary mode that drives the system towards the tetragonal structure in a six atom supercell. b)} A two- dimensional potential-energy surface for ZrO$_2$. The minimum energy structure is set to 0.0 eV and the contour lines correspond to a 0.2 eV increase in energy. The red dot represents a structure corresponding to a high temperature cubic phase of the material. The outsets show the one-dimensional potential-energy surface along each mode.}
    \label{fig:graph_ill}
\end{figure*}

Because the underlying potential-energy surfaces (PES) are complex, relaxing certain polymorphs of a material on these surfaces can be challenging or even impossible.
As an example, zirconia (ZrO$_2$) can exist in its pure form in three different crystal phases: a high temperature ($T > 2370^{ \, \circ}$C) cubic phase, an intermediate temperature ($1170^{ \,\circ}\text{C}\leq T\leq2370^{\,\circ}$C) tetragonal phase, and a low temperature ($T<1170^{ \,\circ}$C) monoclinic phase~\cite{Bocanegra-Bernal2002}.
\REV{In particular, the cubic phase is a dynamically stabilized phase representing an average structure that is rarely, if ever actualized in pristine ZrO$_2$~\cite{Carbogno2014}.
Accordingly, this phase constitutes a saddle point of the PES and its phonon band structure exhibits
an imaginary mode at the X point~\cite{Parlinski1997}, as illustrated in Figure~\ref{fig:graph_ill}a. The associated eigenvector, illustrated in the respective inset,
describes a pairwise, antiparallel distortion of the oxygen atoms that goes hand-in-hand with a stretching of the lattice and leads to the tetragonal structure~\cite{Bocanegra-Bernal2002}.}
To help illustrate this, in Figure~\ref{fig:graph_ill}b we plot a two dimensional PES for a twelve atom zirconia unit cell over a reduced parameter set \REV{describing the cubic lattice parameter $a$ and the motion along the imaginary mode, $z_2$.}
The PES has two wells corresponding to the equivalent tetragonal structures, and a saddle point between them representing a high-symmetry configuration,~i.e.,~the high-temperature cubic phase.
On this PES, a free relaxation of the cubic phase would result in the material relaxing towards a local minima; however, by constraining the relaxation to act only on~$a$, the cubic phase can be obtained as shown in the outsets in Figure~\ref{fig:graph_ill}b.

To help demonstrate how parametric constraints can facilitate addressing these stable/unstable polymorphs, we provide the respective constraints in \REV{Tables \ref{tab:lattice_expressions} and \ref{tab:atomic_expressions}}. As discussed later, these \REV{symmetry blocks} correspond to the input files that are required in the proposed formalism for the twelve atom zirconia unit cell.
For the cubic polymorph, all atoms are at fixed fractional coordinates and only one parameter,~i.e.,~the lattice constant~$a$ can change in a parametrically constrained relaxation.
For the relaxation algorithm, this means that the optimizer can now act only on $a$, all other
degrees of freedom remain untouched. Clearly, this ensures that the initial space group is retained
during relaxation.  As mentioned above, such constraints are necessary, since this polymorph corresponds to a saddle point of the PES, so that an unconstrained or free relaxation would effectively allow the system to break the symmetry and to descend in a local minimum. To explicitly explore these
local minima, the constraints imposed on the cubic cell can be stepwise lifted, whereby the information contained in the imaginary phonon eigenvectors can be incorporated as parametric constraints. As shown in \REV{Tables \ref{tab:lattice_expressions} and \ref{tab:atomic_expressions}}, the pairwise distortion of the oxygen atoms for the imaginary mode at X, cf.~Figure~\ref{fig:graph_ill}a, can be described by introducing one additional parameter for the oxygen distortion,~$z_2$, and one for the tetragonality of the lattice~$c$. These constraints ensure that the geometry optimization occurs along the imaginary phonon mode and leads to the tetragonal minimum. Conversely, a free relaxations can again lead to other local minima of the PES. \REV{In this textbook example, the same constraints could have been imposed by relaxing cubic and tetragonal zirconia in their primitive cells with 6 and 3 atoms, respectively. Generally, this is however not the
case: some materials, e.g., the bismuth oxide~\cite{Matsumoto2010,Drache2007} discussed later, have high-temperature polymorphs with the same number of sites as their stable structures, or more.
}

The previous input example illustrates the flexibility of these constraints, but knowledge of which reduced parameters to use and their relation to the full geometry, must be known before generating an input file.
For crystals, these are determined by the space group and the Wyckoff positions and can therefore be manually constructed.
\REV{Luckily, these parameters are already a part of the definitions used in the AFLOW Library of Crystallographic Prototypes, allowing for an easy way to define these constraints for numerous materials via their utilities~\cite{Mehl2017, Hicks2019}.
The library sorts materials by their space group, stoichiometry, and occupied Wyckoff sites, as calculated with AFLOW-SYM~\cite{Hicks2018b}, placing all materials that share those features into the same crystal prototype~\cite{Mehl2017, Hicks2019}.
A reduced parameter space can then be generated from a prototype definition, and used to describe that class of materials.
For example the tetragonal phase of zirconia can be described by only three parameters: length of the lattice vectors in the $\vec{a}$ and $\vec{b}$ directions ($a$), the ratio of the lattice vectors ($\frac{c}{a}$), and the magnitude of the oxygen distortions ($z_2$).
These parameters represent the same ones we defined earlier from analysis of the phonon bandstructure, with an additional parameter allowing for the relaxation of the lattice upon the atomic distortions.
}
For all prototypes defined in the library, the automatized generation of input geometries for VASP~\cite{Hafner2008}, FHI-aims~\cite{Blum2009}, Quantum Espresso~\cite{Giannozzi2009}, Abinit~\cite{Gonze2002} and more codes is supported by AFLOW.
\REV{The AFLOW library contains 590 unique structure prototypes across all 230 space groups and is thus extremely suitable as a starting point for high-throughput studies.}
As of version 3.1.204, the option \texttt{--add\_equations} can be added to the AFLOW command to generate FHI-aims \texttt{geometry.in} files already containing the additional block required for the constrained relaxation.
Because of this, we use the crystal prototypes defined by the AFLOW Library of Crystallographic Prototypes throughout this work.
\REV{Due to the analytic representation of the parametric expressions it is also straight-forward to add additional parameters to allow for lower symmetric structures or distortions as well as removing parameters to constrain specific components even further.
Additionally, because the AFLOW prototypes are partially based on the space group and occupied Wyckoff sites, it is also straight-forward to adapt their techniques to include structure classes not currently in the library.}

\begin{table*}
\centering
\caption{Parametric expressions for each component of the lattice vectors in the twelve atom cubic and tetragonal ZrO$_2$ supercell}
\label{tab:lattice_expressions}
\begin{tabular}{c|TTT|TTT}
\toprule
\multicolumn{1}{c|}{} & \multicolumn{3}{c|}{Cubic} & \multicolumn{3}{c}{Tetragonal} \\
& $L1^x$ & $L^y$ & $L^z$ & $L^x$ & $L^y$ & $L^z$\\\hline
$L_1$ & \alattice{a} & 0            & 0& \alattice{a} & 0            & 0\\
$L_2$ & 0            & \alattice{a} & 0& 0            & \alattice{a} & 0\\
$L_3$ & 0            & 0 & \alattice{a}& 0            & 0 & \clattice{c}\\
\botrule
\end{tabular}
\end{table*}

\begin{table*}
\centering
\caption{Parametric expressions for each component of the atomic positions in the twelve atom cubic and tetragonal ZrO$_2$ supercell}
\label{tab:atomic_expressions}
\begin{tabular}{c|TTT|TTT}
\toprule
\multicolumn{1}{c|}{} & \multicolumn{3}{c|}{Cubic} & \multicolumn{3}{c}{Tetragonal} \\
Atom                 & $L_1$ & $L_2$ & $L_3$ & $L_1$ & $L_2$ & $L_3$ \\
\hline
Zr & 0.00 & 0.00 & 0.00  & 0.00 & 0.00 & 0.00 \\
Zr & 0.50 & 0.50 & 0.00  & 0.50 & 0.50 & 0.00 \\
Zr & 0.00 & 0.50 & 0.50  & 0.00 & 0.50 & 0.50 \\
Zr & 0.50 & 0.00 & 0.50  & 0.50 & 0.00 & 0.50 \\
O  & 0.25 & 0.25 & 0.25  & 0.25 & 0.25 & 0.25 \zz{- z$_2$} \\
O  & 0.25 & 0.75 & 0.25  & 0.25 & 0.75 & 0.25 \zz{- z$_2$} \\
O  & 0.75 & 0.75 & 0.75  & 0.75 & 0.75 & 0.75 \zz{+ z$_2$} \\
O  & 0.25 & 0.25 & 0.75  & 0.25 & 0.25 & 0.75 \zz{+ z$_2$} \\
O  & 0.75 & 0.25 & 0.25  & 0.75 & 0.25 & 0.25 \zz{- z$_2$} \\
O  & 0.25 & 0.75 & 0.75  & 0.25 & 0.75 & 0.75 \zz{- z$_2$} \\
O  & 0.75 & 0.75 & 0.75  & 0.75 & 0.75 & 0.75 \zz{+ z$_2$} \\
O  & 0.25 & 0.25 & 0.25  & 0.25 & 0.25 & 0.25 \zz{+ z$_2$} \\
\botrule
\end{tabular}
\end{table*}

\REV{In practice, the parametric constraints are implemented in the following fashion.}
Let us assume a $(3\times 3)$-dimensional lattice vector matrix, $\vec{\mathcal{L}}$, and a $(N\times 3)$-dimensional matrix, $\vec{\mathcal{R_F}}$, for the fractional atomic positions.
Given the atomic forces, $\vec{\mathcal{F_R}}$, on the \REV{Cartesian} atomic positions and the stress tensor $\sigma$, we can calculate the derivatives of the energy with respect to the lattice components~\cite{doll}
\begin{align}
	\frac{\rm d E}{\rm d \vec{\mathcal{L}}} = {\vec{\mathcal{L}}^T}^{-1} \,  V \cdot \sigma
\end{align}
where $V$ is the unit cell volume and obtain the generalized forces on the lattice, $\vec{\mathcal{F_L}}$, after cleaning from the atomic contributions
\begin{align}
	\vec{\mathcal{F_L}} = - \frac{\rm d E}{\rm d \vec{\mathcal{L}}}
	- \vec{\mathcal{R_F}}^T \, \vec{\mathcal{F_R}}  .
\end{align}
Each of these matrices denoted by calligraphic letters, $\vec{\mathcal{R_F}}$,  $\vec{\mathcal{F_R}}$, $\vec{\mathcal{L}}$ and $\vec{\mathcal{F_L}}$, can be flattened to one-dimensional vectors that we will name $\vec{R_F}$, $\vec{F_R}$, $\vec{L}$, and $\vec{F_L}$ respectively.
In the parameter representation these quantities reduce to their small-letter counterparts, the $M_R$-dimensional $\vec{r}$, $\vec{F_r}$ and $M_L$-dimensional $\vec{l}$ and $\vec{F_l}$, via
\begin{subequations}
    \begin{align}
    	\label{eq:Rtor}
    	\vec{r} &= \vec{\mathcal{J}_{Rf}}^{-1} \,\left( \vec{R_F} - \vec{t_{Rf}}\right)	\\
    	\label{eq:Ltol}
    	\vec{l} &= \vec{\mathcal{J}_L}^{-1} \,\left( \vec{L}  - \vec{t_L}\right) \\
    	\label{eq:FRtor}
    	\vec{F_r} &= \vec{\mathcal{J}_{R}}^T \, \vec{F_R}	\\
    	\label{eq:FLtol}
    	\vec{F_l} &= \vec{\mathcal{J}_L}^T \, \vec{F_L}	.
    \end{align}
\end{subequations}
where $M_R$ and $M_L$ are the number of free parameters in the atomic and lattice degrees of freedom; \REV{$\vec{\mathcal{J}_{R}}$}, $\vec{\mathcal{J}_{Rf}}$ and $\vec{\mathcal{J}_L}$ are the Jacobian matrix for the transformations and $\vec{t_{Rf}}$ and $\vec{t_L}$ are the \REV{translation} vectors for the respective fractional atomic and lattice degrees of freedom.
\REV{Here $\vec{\mathcal{J}_{R}}$ represents the transformation of the atomic coordinates from Cartesian space to the reduced space, which is calculated from $\vec{\mathcal{J}_{Rf}}$ by
\begin{equation}
    \label{eq:JacobianRr}
    \vec{\mathcal{J}_{R}} = \left( \begin{array}{cccc}{\vec{\mathcal{L}}^T} & {0} & {\dots} & {0} \\ {0} & {\vec{\mathcal{L}}^T} & {\dots} & {0} \\ {\vdots} & {\vdots} & {\ddots} & {\vdots} \\ {0} & {0} & {\dots} & {\vec{\mathcal{L}}^T}\end{array}\right) \vec{\mathcal{J}_{Rf}}.
\end{equation}}
The \REV{translation} vectors are used to include any constant shifts, which are not captured by the Jacobians.
Because \REV{$\vec{\mathcal{J}_{R}}$,} $\vec{\mathcal{J}_{Rf}}$ and $\vec{\mathcal{J_L}}$ are not square and therefore not regularly invertible, we use the generalized left inverse~\cite{rao1972} defined for a matrix $\vec{\mathcal{A}}$ as
\begin{align}
\vec{\mathcal{A}}^{-1,L} = \left( \vec{\mathcal{A}}^T \vec{\mathcal{A}} \right)^{-1} \, \vec{\mathcal{A}}^T ,
\end{align}
provided $\vec{\mathcal{A}}$ has full column rank.
The transformation back to real space can then be performed by inverting Equations \ref{eq:Rtor}-\ref{eq:FLtol}.
The back-transformation of the forces to the full space is not necessary but can be helpful to obtain symmetrized Cartesian or Fractional forces, to check for the convergence of the relaxation.

To facilitate the construction of the Jacobian matrices, we assume a linear relationship between the full coordinates and the parameters.
In principle,$\vec{\mathcal{J}_{Rf}}$ and $\vec{\mathcal{J}_L}$ can be constructed at each step by using analytical expressions to describe each real space degree of freedom as a function of the reduce parameter set; however, by assuming a linear relationship between the spaces they can be initialized at the start of the calculation and used at every step.
For the atomic positions, this assumption is already fulfilled by using fractional instead of Cartesian coordinates.
If we allow angles as unit cell parameters, which is the case for the monoclinic and triclinic lattice systems, the relations become non-linear containing for example expressions like $c\cdot\cos{(\beta)}$.
In these cases, the easiest solution is to substitute each non-zero lattice vector component with an independent parameter.

Before the relaxation the Hessian, $\vec{\mathcal{H}}$, is initialized in the full coordinate space, split into atomic and lattice blocks ($\vec{\mathcal{H}_{R}}$ and $\vec{\mathcal{H}_{L}}$ respectively), and individually transformed into reduced coordinate space, $\vec{\mathcal{H}_{r}}$ and $\vec{\mathcal{H}_{l}}$ via separate Jacobians, $\vec{\mathcal{J}_{R}}$ and $\vec{\mathcal{J}_{L}}$
\begin{align}
    \label{eq:ReduceHessianR}
    \vec{\mathcal{H}_{r}} &= \vec{\mathcal{J}_{R}}^T\vec{\mathcal{H}_{R}}\vec{\mathcal{J}_{R}} \\
    \label{eq:ReduceHessianL}
    \vec{\mathcal{H}_{l}} &= \vec{\mathcal{J}_{L}}^T\vec{\mathcal{H}_{L}}\vec{\mathcal{J}_{L}}.
\end{align}
\REV{Here} $\vec{\mathcal{J}_{R}}$ is \REV{also} divided by the average unit vector length, $V^{1/3}$, so $\vec{\mathcal{H}_{r}}$ and $\vec{\mathcal{H}_{l}}$ are on a similar scale.
The total Hessian is then recombined resulting in
\begin{equation}
    \label{eq:ReducedHessisn}
    \vec{\mathcal{H}} = \left( \begin{array}{cc}{\vec{\mathcal{H}_{r}}} & {0} \\ {0} & {\vec{\mathcal{H}_{l}}}\end{array}\right).
\end{equation}

\begin{figure*}
    \centering
    \includegraphics{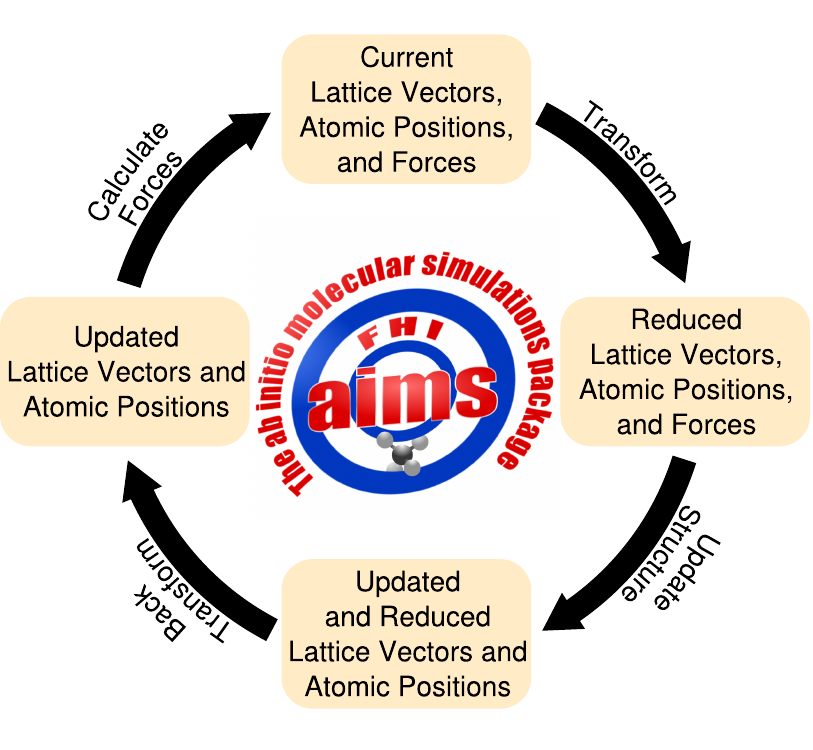}
    \caption{Workflow of the relaxation constrained to the ``parameter reduced space".}
    \label{fig:workflow}
\end{figure*}
Figure~\ref{fig:workflow} illustrates the procedure for relaxing structures with these constraints.
During the relaxation, a full SCF cycle is completed to obtain the forces and the stress tensor for the current geometry, at each step.
If the convergence criterion is fulfilled, i.e. if the forces are below a given threshold, then the relaxation stops and returns the current geometry.
Otherwise the lattice vectors as well as the atomic positions and their respective forces are mapped onto the reduced space using the transformation described in Equations~\ref{eq:Rtor}-\ref{eq:FLtol}.
The atomic coordinates and forces are respectively scaled by $V^{1/3}$ and $V^{-1/3}$, and then passed on to the optimizer. In FHI-aims this is usually a BFGS/TRM optimizer.
Once the optimized parameters are obtained, the full geometry is reconstructed from the parameters and a new relaxation step can begin.

\section{Applications and Benchmarks}
\subsection{Relaxing Metastable and Unstable Systems}
In some cases, constraining a relaxation is necessary to keep the structure in its given polymorph.
Similar to what was seen for zirconia, a material can have many phases that are metastable or unstable at zero point conditions\REV{, but} are stabilized by entropic contributions at higher temperatures or pressures.
Here we define a metastable phase to be one that is in a local minimum on its PES.
While freely relaxing stable or metastable structures is possible by respectively using an initial geometry near its corresponding global or local minimum on the PES, unstable systems will tend to relax towards lower energy and usually lower-symmetric structures, unless they are somehow constrained.

To demonstrate the ability of these constraints to optimize such structures, we relax the twelve atom cubic zirconia unit cell from Figure~\ref{fig:graph_ill}.
While most relaxations will be performed on the primitive cells of structures, we use this system as a simple, demonstrative example.
\REV{The initial structure for the cubic phase was taken from the AFLOW Library of Crystallographic Prototypes, while the initial structure for the tetragonal phase is the same structure with a minor perturbation along the imaginary mode seen in Figure~\ref{fig:graph_ill}.
}
All calculations are done using the FHI-aims package, a full-potential, all-electron electronic structure code.
FHI-aims utilizes numeric atom-centered orbital basis functions, grouped into different tiers beyond the minimal set needed to describe free atoms.
For these calculations we use \textit{tier 1} (double numeric plus polarization basis set) with \textit{light} basis settings which were shown to calculate the lattice parameter and cohesive energy of face-centered cubic gold within 0.001 \AA~ and 20 meV~\cite{Blum2009}, respectively.
We use the PBEsol as the exchange-correlation functional; SCF convergence criteria of $10^{-6}$ eV/\AA~and $5\times10^{-4}$ eV/\AA~for the density and forces, respectively; and the structures are relaxed until the maximum forces on the degrees of freedom are below 0.005 eV/\AA.
All other inputs were taken to be the default values in FHI-aims.
While a larger basis set and using a hybrid functional would increase the accuracy of the calculations, we do not expect it to affect the performance of the relaxation scheme.

\begin{figure*}
	\includegraphics{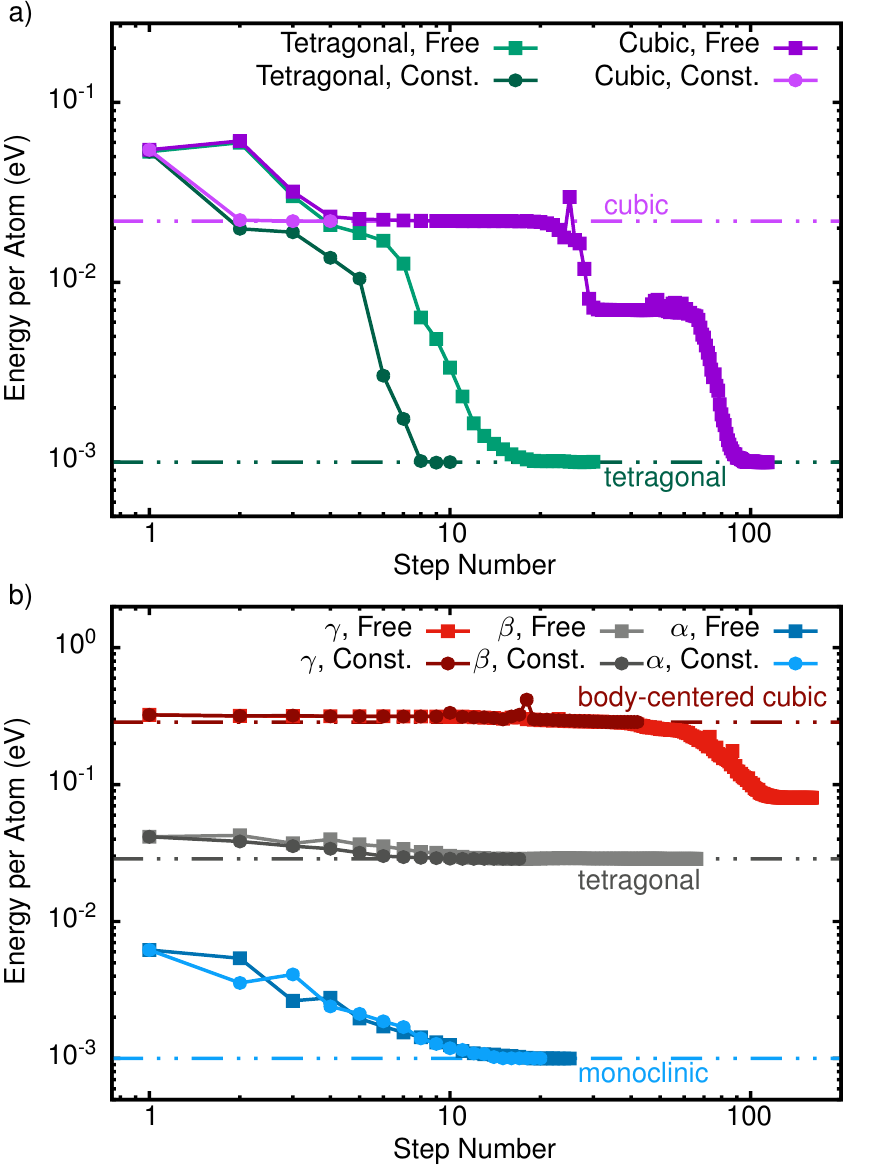}
	\caption{Convergence behaviour of the free (squares) and constrained (circles) relaxations for a) the tetragonal phase (green) and cubic (purple) phases of ZrO$_2$ and b) the \REV{$\alpha$- (blue), $\beta$- (grey) and $\gamma$- (red)} phase of Bi$_2$O$_3$. Both cubic phase of ZrO$_2$ and the \REV{$\gamma$}-phase of Bi$_2$O$_3$ the free relaxation breaks the symmetry and finds an energetically lower structure which is the tetragonal phase for ZrO$_2$ and an unknown phase for Bi$_2$O$_3$. The energy scale on the y-axis is set to 1 meV below the minimum energy for each material.
	}
	\label{fig:convergence}
\end{figure*}

Figure~\ref{fig:convergence}a shows that using the constraints both the cubic and tetragonal phase of ZrO$_2$ can be converged in 4 and 10 steps, respectively, while only the tetragonal phase can be obtained in 30 steps with a free relaxation.
The free relaxation of the cubic phase proceeds towards the tetragonal phase, but initially stalls at a non-physical simple cubic phase in 37 steps.
If the relaxation convergence criteria is further reduced to 0.001 eV/\AA~the structure reaches the tetragonal phase in 114 steps.

Another example of a material with many metastable phases is bismuth oxide.
Bismuth oxide exists in several different polymorphs~\cite{Matsumoto2010} including the low temperature monoclinic phase, $\alpha$-; the high-temperature, face-centered cubic phase, $\delta$-\REV{; the metastable, body-centered cubic phase, $\gamma$-}; and the metastable, tetragonal phase $\beta$-Bi$_2$O$_3$~\cite{Drache2007}.
Upon heating $\alpha$-Bi$_2$O$_3$ transforms into $\delta$-Bi$_2$O$_3$ at around 730$^\circ$C, and remains stable until it melts at approximately 825$^\circ$C~\cite{Drache2007}.
Depending on the cooling procedure, $\delta$-Bi$_2$O$_3$ transitions to \REV{one of the two metastable phases,} $\beta$- \REV{or $\gamma$-Bi$_2$O$_3$,} at \REV{approximately} 650 $^\circ$C \REV{or approximately 640 $^\circ$C, respectively}~\cite{Drache2007}.
Upon further cooling $\beta$-Bi$_2$O$_3$ \REV{and $\gamma$-Bi$_2$O$_3$ respectively} return to the $\alpha$-\REV{phase} at $\sim$300$^\circ$C \REV{and at a temperature dependent on the cooling rate}~\cite{Drache2007}.
Importantly, unlike ZrO$_2$, both the tetragonal \REV{and body-centered cubic phases have as many or more atoms in their} primitive cells \REV{than the $\alpha$ phase}, so \REV{freely relaxing} both structures \REV{in their primitive cells should not prevent them from exploring lower symmetry structures}.
For these calculations we use the same computational settings as those used for ZrO$_2$, but with the \textit{intermediate} settings for the basis set.
The \textit{intermediate} settings and basis sets in FHI-aims increases the accuracy of the default numerical settings, but more importantly adds an $f$-orbital to the double numeric plus polarization basis set for oxygen, which is necessary for describing monoclinic structures containing oxygen.

Figure~\ref{fig:convergence}b shows the relaxation for the \REV{$\alpha$-, $\beta$-, and $\gamma$-} bismuth oxide phases.
\REV{Multiple experimental crystal structure refinements exist~\cite{Drache2007} for both the $\beta$- and $\gamma$- phases, so we take the relaxed structures from the Materials Project, which were initialized from structures taken from the ICSD.}
In \REV{all} cases the constrained relaxation takes less time to converge with the \REV{$\gamma$-,} $\beta$-, and $\alpha$-phase \REV{respectively} needing \REV{42}, 17\REV{,} and 20 steps to converge.
\REV{Freely relaxing both} the $\alpha$ \REV{and $\beta$}-phase\REV{s} simply increases the number of steps needed to converge the system to 25 \REV{and 66} steps\REV{, respectively}, \REV{but removing the constraints for $\gamma$-Bi$_2$O$_3$ causes it to relax to an unknown, non-symmetric structure in 160 steps}.
\REV{While it is possible that the divergent relaxation is a result of an incorrect structure refinement, if that is the case, then it suggests that comparing the final energy and forces of the constrained and free relaxation trajectories can be used to indicate if a structure is correct.}
Since the free relaxation of the  \REV{$\gamma$}-phase departs from the known phases of the material, a more in-depth study of it would be impossible without the constraints as the fully relaxed structures no longer represent the same material.

Both of these cases demonstrate the need for constraining relaxations to their crystal prototype for high-throughput applications.
For the high-symmetry phases of both zirconia and bismuth oxide, the free relaxation not only initially converged to a different phase, but also unknown and potentially physically unrealizable ones.
While the relaxation of ZrO$_2$ does eventually reach one of its known phases, bismuth oxide remains in incorrect structure.
Any further calculations on those structures would be erroneous and could lead to both false positives and skewed property descriptors.
While integrity checks could be made for some of the materials, the consistent breaking of the system's symmetry and the inconsistent degree of that symmetry breaking makes developing standardized checks impractical.
This will be particularly useful for \REV{testing the predictions from} crystal structure discovery \REV{methods} where exact knowledge of lattice type and decorations is necessary.
\REV{Combining these constraints with an automatically generated parametric representation, would provide an efficient means to optimize the newly predicted structures.
}

\subsection{Bench-marking the Algorithm}
\begin{table*}
\centering
\caption{Summary of the materials used in the test dataset}
\label{tab:Material_Dataset}
\begin{tabular}{cccccc}
\toprule
AFLOW Prototype                 & \begin{tabular}[c]{@{}c@{}}Space \\Group \end{tabular} & \begin{tabular}[c]{@{}c@{}}\# of \\Materials \end{tabular} & \begin{tabular}[c]{@{}c@{}}Atoms per\\Unit cell \end{tabular} & \begin{tabular}[c]{@{}c@{}}Free \\Parameters \end{tabular} & \begin{tabular}[c]{@{}c@{}}Full d.o.f. / \\\# Free Parameters \end{tabular}  \\
\hline
AB\_oP8\_62\_c\_c               & 62      & 8           & 8     & 7     & 4.71  \\
A2B\_oP12\_62\_2c\_c            & 62      & 35          & 12    & 9     & 5.00  \\
A2BC4\_tI14\_82\_bc\_a\_g       & 82      & 35          & 7     & 5     & 6.00  \\
A2BC4D\_tI16\_121\_d\_a\_i\_b   & 121     & 29          & 8     & 4     & 8.25  \\
AB2\_hP3\_164\_a\_d             & 164     & 25          & 3     & 3     & 6.00  \\
AB\_hP4\_186\_b\_b              & 186     & 37          & 4     & 2     & 5.25  \\
AB\_cF8\_216\_c\_a              & 216     & \REV{37}    & 2     & 1     & 15.00 \\
ABC\_cF12\_216\_b\_c\_a         & 216     & \REV{54}    & 3     & 1     & 18.00 \\
AB2\_cF12\_225\_a\_c            & 225     & \REV{13}    & 3     & 1     & 18.00 \\
AB2C\_cF16\_225\_a\_c\_b        & 225     & \REV{14}    & 4     & 1     & 21.00 \\
AB\_cF8\_225\_a\_b              & 225     & \REV{19}    & 2     & 1     & 15.00 \\
A\_cF8\_227\_a                  & 227     & 3           & 2     & 1     & 15.00 \\
A2BC4\_cF56\_227\_d\_a\_e       & 227     & 50          & 14    & 2     & 25.50 \\                                                                    \\
\botrule
\end{tabular}
\end{table*}

As the above examples show, the implemented constraints not only ensure that symmetry is preserved, but also accelerate the relaxation because the optimization of a reduced representation with less free parameters is by definition a less demanding task.
To quantify this, the new relaxation algorithm is tested on a set of \REV{359} materials across multiple lattice systems and AFLOW prototypes, as summarized in Table~\ref{tab:Material_Dataset} \REV{and listed in the Supplementary Information}.
\REV{The chosen prototypes represent a sample of common materials with varied space groups and parametric relation complexity.
To further understand how the method performs within a class, we try to include only prototypes with a significant amount of available structures.
}
The initial geometry of each material is taken from either the AFLOW~\cite{Curtarolo2012} or Materials Project~\cite{Jain2013c} database and converted into the right format using the Atomic Simulation Environment~\cite{HjorthLarsen2017}.
All relaxations are done in FHI-aims using the same settings as the zirconia calculations, using both the PBE and PBEsol functionals.

\begin{figure*}
    \centering
    \includegraphics{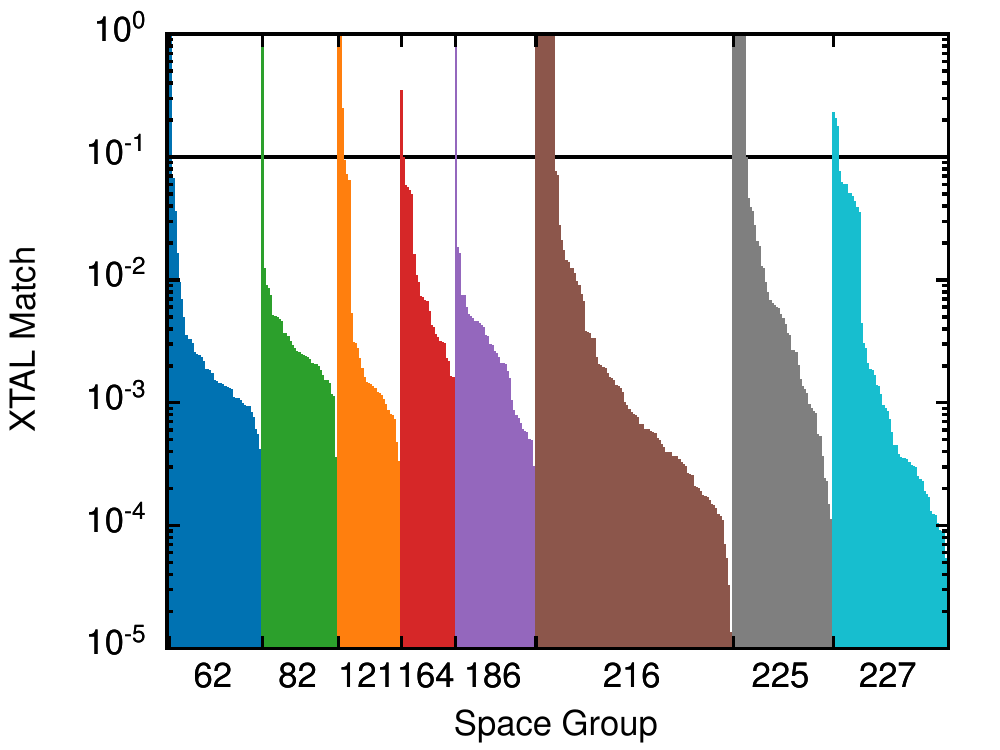}
    \caption{The misfit between the fully and constrained relaxed structures using AFLOW-XTAL-MATCH. All structures below the horizontal line are considered matching. The structures are ordered by space group and then by $m$.}
    \label{fig:relax_xtal}
\end{figure*}

\begin{table*}
\centering
\caption{Summary of the free and constrained relaxation performance by AFLOW prototype.}
\label{tab:relax_sum}
\begin{tabular}{ccc|ccc|ccc}
\toprule

\multicolumn{3}{c|}{} & \multicolumn{3}{c|}{PBE} & \multicolumn{3}{c}{PBEsol} \\
AFLOW Prototype & \begin{tabular}[c]{@{}c@{}}Space\\Group\end{tabular} & \begin{tabular}[c]{@{}c@{}}\# of\\Materials\end{tabular} & \begin{tabular}[c]{@{}c@{}}Average\\Savings\end{tabular} & \begin{tabular}[c]{@{}c@{}c@{}}\# Preserved \\ Space Group \\Free\end{tabular} & \begin{tabular}[c]{@{}c@{}c@{}}\# XTAL \\ Match\end{tabular} & \begin{tabular}[c]{@{}c@{}}Average\\Savings\end{tabular} & \begin{tabular}[c]{@{}c@{}c@{}}\# Preserved \\ Space Group \\Free\end{tabular} & \begin{tabular}[c]{@{}c@{}c@{}} \# XTAL \\ Match\end{tabular} \\

\hline
AB\_oP8\_62\_c\_c               & 62      &      8      &      10.23      &      3        &      8        &      24.61     &      4        &      8       \\
A2B\_oP12\_62\_2c\_c            & 62      &      35     &      10.84      &      19       &      29       &      18.32     &      20       &      33      \\
A2BC4\_tI14\_82\_bc\_a\_g       & 82      &      35     &  \REV{40.32}    &      29       &      32       & \REV{59.98}   &      28       &      34      \\
A2BC4D\_tI16\_121\_d\_a\_i\_b   & 121     &      29     &  \REV{44.01}    &      23       &      26       & \REV{58.13}    &      21       &      26      \\
AB2\_hP3\_164\_a\_d             & 164     &      25     &       7.03      &      10       &      24       &      19.68     &      5        &      24      \\
AB\_hP4\_186\_b\_b              & 186     &      37     &      30.34      &      23       &      36       &      41.68     &      19       &      36      \\
AB\_cF8\_216\_c\_a              & 216     & \REV{37}    & \REV{29.71}     & \REV{28}      & \REV{36}      & \REV{54.83}    & \REV{31}      & \REV{36}     \\
ABC\_cF12\_216\_b\_c\_a         & 216     & \REV{54}    & \REV{33.35}     & \REV{44}      & \REV{54}      & \REV{74.03}    & \REV{36}      & \REV{46}     \\
AB2\_cF12\_225\_a\_c            & 225     & \REV{13}    & \REV{57.63}     & \REV{8 }      & \REV{9 }      & \REV{54.20}    & \REV{7 }      & \REV{9 }     \\
AB2C\_cF16\_225\_a\_c\_b        & 225     & \REV{14}    & \REV{19.77}     & \REV{12}      & \REV{14}      & \REV{80.02}    & \REV{7 }      & \REV{12}     \\
AB\_cF8\_225\_a\_b              & 225     & \REV{19}    & \REV{32.50}     & \REV{11}      & \REV{19}      & \REV{51.72}    & \REV{7 }      & \REV{19}     \\
A\_cF8\_227\_a                  & 227     &      3      &      35.12      &      3        &      3        &      47.62     &      3        &      3       \\
A2BC4\_cF56\_227\_d\_a\_e       & 227     &      50     &      67.10      &      37       &      47       &      73.58     &      37       &      47      \\
\hline
Full Dataset                    &         & \REV{359}   & \REV{34.68}     & \REV{69.64\%} & \REV{93.87\%} & \REV{53.80}    & \REV{62.67\%} & \REV{92.76\%} \\
\botrule
\end{tabular}
\end{table*}

Table~\ref{tab:relax_sum} and Figure~\ref{fig:relax_xtal} illustrate the largest benefit of using the constraints: the preservation of the initial crystallographic prototypes.
Approximately 7\% of all materials tested relax to a different structure according to the AFLOW-XTAL-MATCH tool~\cite{Hicks2019b}.
This tool measures the similarity between two materials using techniques similar to those of Burzlaff, et al.\cite{Burzlaff1997} and produces a misfit value, $m$,
\begin{equation}
    m = 1 - (1 - \text{dev})(1-\text{disp})(1-\text{fail}),
    \label{eq:misfit}
\end{equation}
where dev, disp, and fail are normalized representations of deviations in the lattice vectors, atomic positions, and a failure indicator for when an atomic deviation is more than half the shortest distance in the coordination polyhedra.
From $m$ we can determine if a structure is considered a match using the following mapping
\begin{align*}
	m\leq0.1 :\,\, &\textnormal{structure are similar}	\\
	0.1<m\leq0.2 :\,\, &\textnormal{structures are within same family}	\\
	m>0.2 :\,\, &\textnormal{structures are not compatible}.
\end{align*}
All \REV{of} the \REV{26} materials with divergent relaxations in the data set follow the same pattern as cubic-ZrO$_2$ and \REV{$\gamma$}-Bi$_2$O$_3$.
Therefore constraining these material's relaxation is vital because without them all further calculations on these materials would no longer be physically relevant.

Constraining the relaxation has benefits even when the final structures are similar as it significantly reduces the number of steps needed for the trajectories to converge.
When the constrained and free relaxations proceed towards the same structures, the constraints reduce the number of relaxation steps by an average of \REV{33.11}\% and \REV{52.43}\% for calculations using the PBE and PBEsol functional, respectively.
Including the divergent structures respectively increases the saving to \REV{34.68}\% and \REV{53.80}\%, but this is expected as the constrained relaxation and free relaxations are acting on qualitatively different PESs.
Here we define the savings, $S$,  as
\begin{equation}
    S = \frac{N_\text{free}-N_\text{constrained}}{N_\text{constrained}} \times 100\%
\end{equation}
where $N_\text{free}$ is the number of steps need to converge the free relaxation and $N_\text{constrained}$ is the number of steps need to converge the constrained relaxation.
The seemingly better performance of the constraints when using the PBEsol, is likely because of larger differences between the initial and final structures when using this functional.
On average comparing the starting geometries with the freely relaxed ones gave an average $m$ of \REV{0.07654} (\REV{0.01630} when $m\neq1$) and \REV{0.1322} (\REV{0.02038} when $m\neq1$) for the PBE and PBEsol calculations, respectively.
And since increasing the distance to the final structure also increases the likelihood of the free relaxation deviating from the constrained trajectory, larger savings using the PBEsol functional is expected.
This is also supported by the larger differences between the final structures of constrained and free relaxations for the PBEsol functional.

\begin{figure*}
    \centering
    \includegraphics{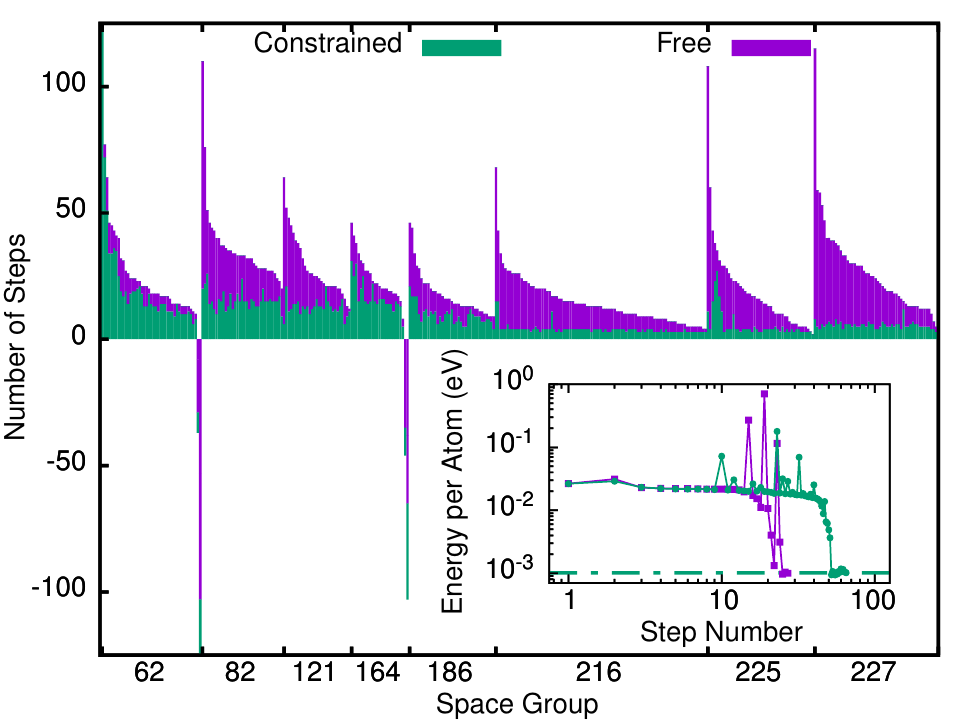}
    \caption{
        The total number of steps needed for the constrained (green) and free (purple) relaxations. Negative step numbers represent the cases where the constrained relaxations take longer than the free relaxation. The bar for orthorhombic N$_2$O is not shown in the figure, the free relaxation took 310 steps, and the constrained one required 296 steps to converge. The inset shows the constrained (green circles) and free relaxation (purple squares) trajectory of platinum (IV) sulfide, which is one of the materials where the constrained relaxation takes more steps than the free relaxation. The zero of the energy scale is set to 1~meV below the energy of the relaxed structure (horizontal dashed line).
    }
    \label{fig:relax_steps}
\end{figure*}

Unfortunately, the savings are not consistent across the various prototypes.
Figure~\ref{fig:relax_steps} shows the total number of steps needed to relax the structures with and without constraints for the PBEsol calculations, sorted by the space group and then the maximum number of relaxation steps needed.
Similar to the average savings shown in Table~\ref{tab:relax_sum}, as the number of free parameters approaches the number of degrees of freedom in a material the savings from constraining the relaxation decrease.
In some cases using the constraints actually increase the number of steps needed for the relaxation to converge, but in all of these cases \REV{for PBEsol} the relaxation trajectories take unproductive steps shown in the inset of Figure~\ref{fig:relax_steps} for PtS$_2$.
In these cases the extra \REV{degrees of freedom} allow the relaxation trajectories to pass the problematic regions faster and therefore converge to the final structures in fewer steps.
\REV{For the PBE calculations, there are also some cases where the relaxation takes a few extra steps at the end of the trajectory where the total forces are near convergence criteria.}
These results suggest that for lower symmetry structures, the potential savings from constraining the relaxation can be considerably smaller.

Constraining the relaxation can also decrease the computational time needed to perform further calculations after the relaxations.
Despite their similarity to the final geometries of the constrained relaxation, freely relaxed structures always break symmetry to some degree.
To measure how well the relaxation preserves symmetry we compare the \textit{spglib} calculated space groups of the initial and converged structures~\cite{Togo2018}.
\textit{spglib} calculates the space group of a material by iteratively searching for a given structure's primitive cell and symmetry operations and using those to generate the space group.
The algorithm determines if a structure's space group has a certain symmetry operation, by checking if the operator transforms all the atoms in the structure to sites occupied by the same type of atom within a given small euclidean distance, $\varepsilon$.
By default $\varepsilon$ is set to 10$^{-5}$ \AA, and none of the tested materials remain in their initial space groups using this setting.
To preserve the symmetry for the freely relaxed structures, a larger user-defined tolerance threshold is needed (0.01 \AA~is used in Table II), consistent with findings reported by Hicks, \textit{et al.}~\cite{Hicks2018b}.
However, when using the constraints, the symmetry is always perfectly preserved for all materials.
This is advantageous as exploiting symmetry can greatly reduce the time needed to do further calculations for many applications.
In particular phonon calculations using the finite difference approach where the symmetry of the system determines the number of atomic displacements, and therefore the number of force calculations, needed to calculate a complete set of force constants.
By default \textit{phonopy}, a very popular Python package for calculating phonon spectra with finite differences~\cite{phonopy}, uses the default \textit{spglib} settings to calculate the symmetry of a material, which would misrepresent all of the tested materials using free relaxation.
Commensurate phonon calculations can be achieved by performing symmetry constrained relaxations and using more robust symmetry tools, such as AFLOW-SYM~\cite{Hicks2018b}.

\subsection{Systems with local symmetries or distortions}
\label{sec:polaron}
One of the key advantages of defining constraints in this way is the ability to locally break symmetry to account for point defects.
While the previously discussed benefits of lower relaxation times and relaxing unstable structures could also be achieved by using symmetrized forces, that method would not be able to locally break its symmetry.
However many of these defects exhibit some type of short range order, such as Jahn-Teller-type lattice distortions~\cite{Rodriguez2018,Prentice2017,Evarestov2012}, that can be parametrically added on top of the standard crystal structure.
By including these distortions one can reduce the computational cost of relaxing supercells with defects, which is important when using a supercell approach to study point defects.
This approach uses density functional methods to calculate the total energy of a system in a series of supercells of increasing size to study the effects of a defect on the system.
If a scaling law is known for the system the results of these calculation can be extrapolated to the experimentally relevant dilute limit.
As the supercell size increases the time needed to calculate each step also increases, therefore any reduction in the number of steps needed can save a significant amount of computational time.

\begin{figure*}
	\includegraphics{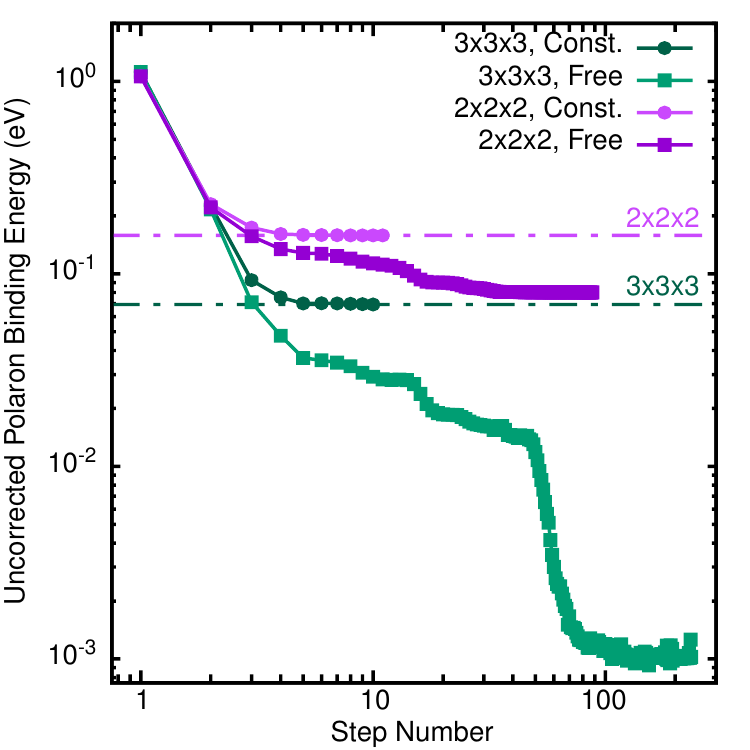}
	\caption{The convergence of the free (squares) and constrained (circles) relaxation of a polaronic distortion in a 2x2x2 (purple) and 3x3x3 (green) supercell of rock-salt MgO. The zero of energy scale on is set to the energy of the undistorted lattice. The horizontal dashed lines correspond to the final energy of constrained relaxation.
	}
	\label{fig:local_sym_breaking_convergence}
\end{figure*}
To illustrate the ability of the new relaxation scheme to study point defects in materials, we study a polaronic distortion in MgO previously studied in our department \REV{at the Fritz Haber Institute}~\cite{Kokott2018}.
Polarons are quasiparticles that couple point charges with lattice distortions in a material, reducing its charge carrier mobility.
\REV{
Typically, the size of the polaron is controlled by the electron-phonon interaction, with a stronger interaction leading to a smaller polaron~\cite{Kokott2018,Nag2014}.
Because of the reduced charge carrier mobility, understanding how polarons form and migrate through a material is vital for a number of applications ranging from catalysis~\cite{Dohnalek2010,Henderson2011,Rettie2016} to thermoelectricity~\cite{Nag2014}.}
For rock salt MgO the lattice distortions for an electron hole polaron are Jahn-Teller like\REV{, with the oxygen and magnesium ions being respectively attracted and repelled from the hole.
To model this system, we assume the electron hole is located on a fixed oxygen atom placed in the center of the supercell.
We then allow the other ions to relax from their initial position along a line going through the center of the cell, with the magnitude and sign of the motion being controlled by a single free parameter for each atom.
This constraint choice reduces the number of degrees of freedom to $N-1$, where $N$ is number of atoms in the supercell.
While these constraints do not account for periodic images of the electron hole or changes in the positions of the other ions, it does preserve the main polaronic distortion.
Moreover, the restrictiveness of the constraints can be decreased further on subsequent calculations until the desired level of accuracy is achieved.
}

Figure~\ref{fig:local_sym_breaking_convergence} illustrates the uncorrected polaron energy for the free and constrained relaxation trajectories for these supercells~\cite{Kokott2018}.
Because charge localization is necessary when studying polarons, the HSE 2006 functional is used with a screening parameter of 0.11 Bohr$^{-1}$ with exact exchange.
\REV{For the unrestricted spin relaxation} the SCF density, forces, total energy, and eigenvalues were all converged to 10$^{-4}$ eV/\AA, 10$^{-4}$ eV/\AA, 10$^{-5}$ eV, and 10$^{-2}$ eV, respectively.
Each atom calculated 5 empty states above those used by the Kohn-Sham orbitals and the structures were relaxed until the total forces on the free parameters were below 10$^{-4}$ eV/\AA.
\REV{The constrained relaxation for the distorted geometry converges the 64 atom supercell in 11 steps which is only one-eighth of the steps needed by the free relaxation.
Significantly better performance is seen for the 216 atom supercell, which converges in 10 constrained relaxation steps, 96\% less than in the unconstrained case needing 234 steps.
This increase in efficiency is likely a result of the constraints focusing on the primary Coloumbic interaction, reducing the amount of fine-tuning it needs to do to balance out the weaker perturbation in the other inter-ionic interactions.
}
Using the constraints does cause the relaxation to converge to a structure with slightly higher energy, but this is because the chosen constraints are valid for an isolated polaron and thus do not account for the artificial interactions with periodic images due to the finite supercell size.
In practice, this effect is rather negligible, since we find that the obtained polaron energy is 78.4~meV higher in energy than the free relaxation in the 2x2x2 supercell and 69.1~meV higher in the 3x3x3 supercell.

\section{Conclusions and Outlook}
In this work we presented a new scheme for parametrically relaxing structures in a \REV{general,} symmetry-reduced space.
After explaining the algorithm, we test it on \REV{359} different materials across a broad range of material classes.
In all cases the new method was able to strictly preserve the symmetry of the materials, and on average reduced the number of steps needed to converge a material by 50\%.
We also demonstrated \REV{for the example of bismuth oxide} how the constraints can be used to relax to metastable phases. \REV{Finally, we showcased the relaxation of structures with local symmetry breaking with known distortion patterns for polarons in MgO.}

This new method will have a profound impact on computational materials discovery.
Not only does the decreased cost of relaxing a material increase the velocity of high-throughput search, but it also allows for those searches to explore metastable and dynamically stabilized structures.
The method also has the promise to improve the efficiency of supercell calculations and study only the physically relevant structures.
Finally by monitoring the difference between the full forces and symmetrized forces new stable phases can potentially be discovered from metastable or unstable polymorphs.
Although we showed that the proposed algorithm is applicable to accelerate and improve standard solid-state physics calculations, its flexibility allows it to be applied to a much wider range of problems, e.g., transition-state searches \REV{or interface relaxations}.
Similarly, it is easily generalizable to other form of coordinates and straightforwardly implementable in any electronic-structure theory code\REV{, and has already been implemented as constraints within ASE}.

\subsection{Data Availability}
The full input and output files for the calculations in this work are available in the NOMAD repository with the identifier [DOI:10.17172/NOMAD/2019.10.19-1]~\cite{Lenz2019}.

\subsection{Code Availability}
The algorithms presented in this paper have been fully implemented in FHI-aims and is available in any version after 20191019. The constraints have also been implemented in ASE\url{https://gitlab.com/ase/ase/merge_requests/1298}.

\subsection{Author Contributions}
MOL and TARP contributed equally on this work by implementing and bench-marking the discussed relaxation scheme. DH worked to incorporate the constraints into AFLOW. SC, MS, and CC directed the project. All authors analysed the data and wrote the manuscript.

\subsection{Acknowledgements}
TARP would like to acknowledge Florian Knoop for valuable discussions.
\REV{The authors also would like to Acknowledge reviewer 3 for pointing out different refinements of the metastable Bi$_2$O$_3$ phase.}
This project was supported by TEC1p (the European Research Council (ERC) Horizon 2020 research and innovation programme, grant agreement No 740233), BigMax (the Max Planck Society's Research Network on Big-Data-Driven Materials-Science), and the NOMAD pillar of the FAIR-DI e.V. association.
SC and DH acknowledges U.S. DOD-ONR (Grants No. N00014-17-1-2090). D.H. acknowledges support from the U.S. DOD through the National Defense Science and Engineering Graduate (NDSEG) Fellowship Program.
We thank the Max Planck Computing and Data Facility for computational resources.

\bibliography{bibliography}
\bibliographystyle{unsrt}

\end{document}